\documentclass[aps, 10pt]{revtex4}
\usepackage{graphicx,subfigure}
\usepackage{epsfig}
\usepackage{amsmath}
\usepackage{amssymb}
\usepackage{amsthm}
\usepackage{bbold}
\begin{document}
\title{Quantum Hydrodynamics: Kirchhoff Equations }
\author{ K. V. S. Shiv Chaitanya }
\email[]{ chaitanya@hyderabad.bits-pilani.ac.in}
\affiliation{Department of Physics, BITS Pilani, Hyderabad Campus, Jawahar Nagar, Shamirpet Mandal,
	Hyderabad, India 500 078.}

\begin{abstract}
In this paper, we show that the Kirchhoff equations are derived from  the Schr\"odinger  equation by assuming the wave function to be a polynomial like solution. These Kirchhoff equations  describe the evolution of $n$ point vortices in hydrodynamics. 
In two dimensions,  Kirchhoff equations are used to demonstrate the solution to single particle Laughlin wave function as complex Hermite polynomials. We also show that the equation for optical vortices, a two dimentional system, is derived from Kirchhoff equation by using  paraxial wave approximation.  These Kirchhoff equations  satisfy  a Poisson bracket relationship in phase space which is identical to the Heisenberg uncertainty relationship. Therefore, we conclude that being classical equations,  the Kirchhoff equations,  describe both a particle and a wave  nature of single particle quantum mechanics in two dimensions. 
\end{abstract}

\maketitle

\section{Introduction}

Nine different formulations of non-relativistic quantum mechanics exist \cite{nine}. They are,
wavefunction formalism, matrix mechanics, path integral formalism, phase space formalism, density matrix formalism, second quantization, variational formalism, pilot wave theory, and Quantum Hamilton-Jacobi (QHJ) formulations. Of these, the wave function formalism and the matrix mechanics are popular and  few other formalisms  attempt to map the quantum mechanics to classical mechanics. The well known one's are the phase space formalism \cite{wig} and pilot wave theory. In the phase space formulation, where Wigner quasi-probability distribution \cite{wig} is defined such that it links the wave function that appears in Schr\"odinger  equation  to a probability distribution in phase space, but this formalism has a draw back of  negative probabilities. This mapping of quantum mechanics to classical mechanics is of central importance to the philosophy of physics, and also the interpretation of quantum mechanics.

One of the first attempts to find a classical relationship is given by  Erwin Madelung known as Madelung quantum hydrodynamics \cite{med} and 
the Madelung equations 
\begin{equation}
\partial _{t}\rho _{m}+\nabla \cdot (\rho _{m}{\mathbf {u} })=0,\label{e1}
\end{equation} 
and \begin{equation}
{\displaystyle {\frac {d{\mathbf {u} }}{dt}}=\partial _{t}{\mathbf {u} }+{\mathbf {u} }\cdot \nabla {\mathbf {u} }=-{\frac {1}{m}}\mathbf {\nabla } \left(Q+V\right)}\label{e2}
\end{equation} 
where ${\displaystyle {\mathbf {u}}} $ is the flow velocity, 
$ {\displaystyle \rho _{m}=m\rho =m|\psi |^{2}} $ is the mass density, $ {\displaystyle Q=-{\frac {\hbar ^{2}}{2m}}{\frac {\nabla ^{2}{\sqrt {\rho }}}{\sqrt {\rho }}}=-{\frac {\hbar ^{2}}{2m}}{\frac {\nabla ^{2}{\sqrt {\rho _{m}}}}{\sqrt {\rho _{m}}}}} $ is the Bohm quantum potential, and $V$ is the potential from the Schr\"odinger equation. The Kirchhoff equations are given by
\begin{equation}
\frac{d \bar{z}_i}{dt}=\sum_{1\leq i\leq n,i\neq j}^n\frac{i\Gamma_i}{z_i-z_j}+ iW(z_i)\label{pt1}
\end{equation}
which describe the evolution of $n$ point vortices in incomprehensible fluid \cite{aref}, where 
$z_i=x_i+iy_i$ are position of vortices, $\Gamma_i$ the circulation strength, $W(z_i)$ is background flow and 
$\bar{z_i}=x_i-iy_i$.  are derived from Euler equations (\ref{e1}) and (\ref{e2}), for the complex velocity \cite{mast}.
Readers should note that in deriving Kirchhoff equations quantum potential $Q$ is not considered. 

This  Madelung quantum hydrodynamics formalism was later modified by Bohm, known as the pilot wave formulation \cite{bo,bo1}.  de Broglie first proposed the pilot wave theory of double solution \cite{de}, to explain wave particle duality. In this theory, the Schr\"odinger equation has two solutions, one the regular wave function $\psi(x,y,z)=ae^{i\phi(x,y,z)}$ where $a$ is a constant, and the other physical wave solution $u(x,y,z)=f(x,y,z)e^{i\phi(x,y,z)}$. The two solutions are related by the phase $\phi(x,y,z)$. The singularities in $u(x,y,z)$ are due to presence of singularities in $f(x,y,z)$ give rise to particle-like nature. These are  moving singularities. The drawback of the theory is that it explains only single particle case. This theory was further developed by Bohm \cite{bo,bo1} for a system of many particles. In this theory, the wave particle duality vanishes, quantum system behaves like a particle and a wave, simultaneously, in the same experimental setup. Then the particles are directed  by the pilot wave which will guide them to areas of interference. 

In this paper, the Kirchhoff equations are derived from  the Schr\"odinger  equation by assuming the wave function to be a polynomial like solution. These Kirchhoff equations  describe the evolution of $n$ point vortices in hydrodynamics.  These Kirchhoff equations are classical equations of motion. Therefore, classical mechanics structure of the Kirchhoff equations admits both  particle nature and wave nature of single particle quantum mechanics in two dimensions.

In literature, the fractional quantum Hall effect ground state evolution is described by Laughlin wave function \cite{rbl}, which is an ansatz,  modeled in terms of Kirchhoff equations  \cite{pb1,pb2,pb3,aga}. 
These Kirchhoff equations (\ref{pt1}) are in phase space and admit a Poisson bracket relationship \cite{aref}
in terms of the complex coordinates given by
\begin{equation}
\{z_i,\bar{z}_j\}=-2i\delta_{ij}\Gamma_i.\label{ps}
\end{equation} 
For more details, refer to \cite{aref}. The  identification of $\bar{z}=\partial_{z_i}=P_z$ with canonical momenta \cite{pb1} allows one to  replace Poisson bracket with commutator 
\begin{equation}
[z_i,\partial_{z_i}]=i\hbar,\label{hs}
\end{equation}  
where $\Gamma_i=\hbar$ is the  Heisenberg uncertainty relation.

 As an illustration we  show that the logarithmic derivative of the Laughlin wave function looks like the right-hand-side of the Kirchhoff equation and their one particle solutions are complex
Hermite polynomials. Then, we address the wave particle duality through Kirchhoff equation in terms of interference or as a relative phase. We show that the equation for optical vortices is derived from Kirchhoff equation using paraxial wave equation in presence of  real constant background.

\section{Schr\"odinger  equation and Kirchhoff equations}
In this section, the Kirchhoff equations are derived from  the Schr\"odinger  equation by assuming the wave function to be a polynomial like solution  \cite{mast}. These equations describe the evolution of $n$ point vortices in Hydrodynamics.\\
Consider the time dependent Schr\"odinger  equation with potential $V(x)=0$ 
\begin{equation}
i {\frac {\partial }{\partial t}}\psi (x ,t)=\Gamma\frac{\partial^2}{\partial x^2}\psi (x ,t),\label{ts}
\end{equation}
where $\Gamma={\frac {-\hbar }{2m }}$.
By introducing a polynomial
\begin{eqnarray}
\psi(x,t)&=&(x-x_1(t))(x-x_2(t))\cdots (x-x_n(t))\nonumber\\&=&\prod_{k=1}^n(x-x_k(t)),\label{poly}
\end{eqnarray}
for $n=2$ substituting equation (\ref{poly}) in  equation (\ref{ts}) 
\begin{equation}
-i\dot{x}_1(x-x_2(t))-i\dot{x}_2(x-x_1(t))=2\Gamma
\end{equation}
The equations at $x=x_1$ and $x=x_2$
\begin{equation}
\dot{x}_1=\frac{2\Gamma i}{(x_1-x_2)},\;\;\;\;
\dot{x}_2=\frac{2\Gamma i}{(x_2-x_1)}.
\end{equation}
By following a similar procedure for $n=3$, the equations at points $x=x_1$, $x=x_2$ and $x=x_3$ are given by
\begin{eqnarray}
\dot{x}_1&=&2\Gamma i\left[\frac{1}{(x_1-x_2)}+\frac{1}{(x_1-x_3)}\right]\nonumber\\
\dot{x}_2&=&2\Gamma i\left[\frac{1}{(x_2-x_1)}+\frac{1}{(x_2-x_3)}\right]\nonumber\\
\dot{x}_3&=&2\Gamma i\left[\frac{1}{(x_3-x_1)}+\frac{1}{(x_3-x_2)}\right]\nonumber
\end{eqnarray}
The same procedure for $n$ zeros gives
\begin{eqnarray}
\dot{x}_i=2\Gamma i\sum_{i\neq j}^n\frac{1}{(x_i-x_j)}.\label{kir1}
\end{eqnarray}
The  equation (\ref{kir1}) are known as Kirchhoff equations which describe the evolution of $n$ point vortices in Hydrodynamics.
Therefore, Schr\"odinger equation can be written as a system of $n$ linear equations (\ref{kir1}). Kirchhoff equations (\ref{kir1}) with the background flow
$\mathcal{W}(x)$ are given by
\begin{eqnarray}
\dot{x}_i=2\Gamma i\sum_{i\neq j}^n\frac{1}{(x_i-x_j)} +i\mathcal{W}(x_i).\label{kir}
\end{eqnarray}
Solution to the stationary Kirchhoff equations (\ref{kir}), that is $\dot{x}_i=0$, is found by Stieltjes electrostatic model \cite{st,st1}. In this model, there are $n$ unit moving charges between two fixed charges $p$ and $q$ at $-1$ and $1$ respectively, on a real line and it is shown by Stieltjes that the system attains equilibrium at the zeros of Jacobi polynomials. Further, it is proved by the author that, Stieltjes electrostatic model is analogous to the quantum momentum function of  quantum Hamilton Jacobi \cite{kvs}. Here the moving unit charges are replaced by moving poles. They  are similar to imaginary charges with $i\hbar$ placed between two fixed poles like fixed charges.  In the process, the background flow $\mathcal{W}(x)$ is identified with the superpotential.

In supersymmetry, the superpotential $\mathcal{W}(x)$ is defined in terms of intertwining operators $ \hat{A}$ and $\hat{A}^{\dagger}$  as
\begin{equation}
\hat{A} = \frac{d}{dx} + \mathcal{W}(x), \qquad \hat{A}^{\dagger} = - \frac{d}{dx} + \mathcal{W}(x).
\label{eq:A}
\end{equation}
from this a pair of factorized Hamiltonians $H^{\pm}$ is defined as
\begin{eqnarray}
H^{+} &=& 	\hat{A}^{\dagger} \hat{A} 	= - \frac{d^2}{dx^2} + V^{+}(x) - E, \label{vp}\\
H^{-} &=& 	\hat{A}  {\hat A}^{\dagger} 	= - \frac{d^2}{dx^2} + V^{-}(x) - E, \label{vm}
\end{eqnarray}
where $E$ is the factorization energy. The partner potentials $V^{\pm}(x)$ are related to $\mathcal{W}(x)$ by 
\begin{equation}
V^{\pm}(x) = \mathcal{W}^2(x) \mp \mathcal{W}'(x) + E, \label{gh}
\end{equation}
where  prime denotes differentiation with respect to $x$.

As an illustration, we solve the Harmonic oscillator  in natural units using the Kirchhoff equations 
\begin{equation}
\sum_{1\leq j\leq n,j\neq k}^n\frac{1}{x_k-x_j}- x_j=0,\label{ufip}
\end{equation}
where $\mathcal{W}(x)=x_j$  is the superpotential of  Harmonic oscillator.\\ 
By introducing a polynomial
\begin{eqnarray}
f(x)=(x-x_1)(x-x_2)\cdots (x-x_n),\label{poly1}
\end{eqnarray}
and taking the  limit $x\rightarrow x_j$ and using l'Hospital rule we obtain
\begin{eqnarray}
\sum_{1\leq j\leq n,j\neq k} \frac{1}{x_j-x_k}&=&\lim_{x\rightarrow x_j}
\left[\frac{f'(x)}{f(x)}-\frac{1}{x-x_j}\right]\nonumber\\&=&\lim_{x\rightarrow x_j}\frac{(x-x_j)f'(x)-f(x)}{(x-x_j)f(x)}
\nonumber \\&=& \frac{f''(x_j)}{2f'(x_j)}.\label{id}
\end{eqnarray}
By substituting equation (\ref{id}) in equation (\ref{ufip}), we obtain 
\begin{equation}
f''(x_j)+2 x_jf'(x_j)=0.\label{poky}
\end{equation}
Hence equation (\ref{poky}) is a polynomial of order $n$, and is proportional to f(x) which gives Hermite differential equation
\begin{equation}
f''(x)+ 2xf'(x)+nf(x)=0.\label{pokyj}
\end{equation}
Here it should be noted that when we solve the problem for general potential say $Q(x_j)$ through Stieltjes electrostatic model  we end up with the following polynomial solutions
\begin{equation}
f''(x_j)+Q(x_j)f'(x_j)=0.\label{pokylu}
\end{equation}
Thus, the differential equation (\ref{pokylu}) will have a classical orthogonal polynomial solution only  when $Q(x_j)=\mathcal{W}(x_j)$.  

As an illustration, consider the Coulomb potential whose superpotential is given by 
\begin{equation}
\mathcal{W}_{coul}(r_j)=\frac{1}{2}-\frac{(l+1)}{r_j}.
\end{equation}
Then the  Kirchhoff equations 
\begin{equation}
\sum_{1\leq j\leq n,j\neq k}^n\frac{1}{r_k-r_j}- _=\frac{1}{2}-\frac{(l+1)}{r_j}=0,\label{ufipc}.
\end{equation}
Using the identity (\ref{id}) and substituting in Kirchhoff equation (\ref{ufipc}) gives
\begin{equation}
\frac{f''(r_j)}{2f'(r_j)}-(\frac{1}{2}-\frac{(l+1)}{r_j})=0.
\label{cop}
\end{equation}
The equation (\ref{cop}) is polynomial of order $n$ given by
\begin{equation}
rf''(r)+(2(l+1)-r)f'(r)=0,
\end{equation}
proportional to $f(r)$  gives the Laguerre  differential equation
\begin{equation}
rf''(r)+(2(l+1)-r)f'(r)+nf(r)=0.
\end{equation}
Similarly, for super potential
\begin{equation}
\mathcal{W}(x_j)=\frac{p}{x_j-1}+\frac{q}{x_j+1}.\label{jac},
\end{equation} 
the  Kirchhoff equations are given by
\begin{equation}
\sum_{1\leq j\leq n,j\neq k}^n\frac{1}{x_k-x_j}- \frac{p}{x_j-1}-\frac{q}{x_j+1}=0.\label{ufipj}
\end{equation}
Using the identity (\ref{id}) and substituting in Kirchhoff equation (\ref{ufipj}) gives
\begin{equation}
-\frac{f''(x_j)}{2f'(x_j)}-\frac{p}{x_j-1}-\frac{q}{x_j+1}=0.\label{pok}
\end{equation}
The equation (\ref{pok}) is polynomial of order $n$ given by
\begin{eqnarray}
(1-x^2)f''(x)+2[q - p - (p + q)x]f'(x)=0.\label{poi}
\end{eqnarray} 
is proportional to $f(x)$  gives the Jacobi differential equation
\begin{eqnarray}
(1-x^2)f''(x)+2[q - p - (p + q)x]f'(x)+n(n+p+q+1)f'(x)=0.
\end{eqnarray} 
Therefore, we can solve all the bound state problems in terms of Kirchhoff equations  by  using the Stieltjes electrostatic model.

\section{Laughlin Wave Function}
One of the well known examples of two dimensional systems is fractional quantum Hall effect. Robert Laughlin proposed the following wave function as an ansatz for the ground state wave function of fractional quantum Hall effect \cite{rbl}
\begin{equation}\label{lf}
\psi =  \prod_{N \geq j > i \geq 1}\left( z_j-z_i \right)^n  \exp\left( - \frac{1}{4l_B^2} \sum_i\vert z_j\vert ^2 \right)
\end{equation}
where, $z_i=x_i+iy_i$,  $N$ is number of electrons, $l_B=\sqrt{\frac{\hbar}{eB}}$ is magnetic length, $\hbar$ is Planck constant, $e$ is electric charge, $B$ is magnetic field, $\omega_B$ is cyclotron frequency $\omega_B=\frac{eB}{m}$, $z_i$ and $z_j$ are the position of electrons for the ground state of a two-dimensional electron gas with the lowest Landau level, where $n$ is written in terms of $\nu=1/n$ and  $n$ is an odd positive integer filling numbers. As the Laughlin wave function is a trial wave function, it is not an exact ground state of any potential, in particular, not an exact ground state of Coulomb repulsion problem. But it has been tested numerically, for the Coulomb and several repulsive potentials, that the Laughlin wave function has more than $99\%$ overlap with the true ground state \cite{dt}. The gap vanishes at the edges and the fractional charges are calculated using Berry's connection. In literature, several model Hamiltonians have been proposed which can admit Laughlin wave function as a solution and first model of its kind was proposed by Haldane \cite{hal}.
The Berry connection for $N$ quasi-holes are given by \cite{dt}
\begin{equation}  
A(\eta_j)=-\frac{i\nu}{2}\sum_{1\leq k\leq N,j\neq k}\frac{1}{\eta_k-\eta_j}+i\nu\frac{\bar{\eta_j}}{4l_B^2},\label{b1}
\end{equation}
here $\eta_i$ are the positions of the quasi holes and the $\nu$ are the filling numbers defined in equation (\ref{lf}). A similar equation for $A(\bar{z_j})$ exist which describes the adiabatic transport of quasihole  at $z_i$ when all other quasihole positions are fixed. 
Taking logarithm of Laughlin wave function (\ref{lf}) and then differentiating  with respect to $z$, equating the derivative to zero gives Kirchhoff equations.
\begin{eqnarray}
i\frac{d}{dz_j}ln \psi(z_j)=\sum_{1\leq i\leq N,i\neq j}\frac{in}{z_i-z_j}-i\frac{1}{4l_B^2}\bar{z}_j =0.\label{kri}
\end{eqnarray}
Equation (\ref{kri}) is  Berry's Connection (\ref{b1}) for the positions of quasi hole,s and the $\nu$'s are filling numbers defined in equation (\ref{lf}). Similarly, by taking the transpose of  equation (\ref{kri}) we obtain an equation for $A(\bar{z_j})$
\begin{eqnarray}
\sum_{1\leq i\leq N,i\neq j}\frac{in}{\bar{z}_i-\bar{z}_j}-i\frac{1}{4l_B^2}z_j =0.\label{kri1}
\end{eqnarray}
which describes the adiabatic transport of quasihole  at $z_i$ when all other quasihole positions are fixed.
Therefore, from equation (\ref{kri}),  it is clear that the Laughlin wave function (\ref{lf}) represents a Hamiltonian, and the Kirchhoff equations are obtained by taking the minimum  of logarithm of the Laughlin wave function  (\ref{lf}). This represents  time independent Schr\"odinger  equation  with background flow $W(\bar{z_j})=\frac{1}{4l_B^2}\bar{z_j}$.

The equations (\ref{kri}) and  (\ref{kri1}) are Schr\"odinger  equations, and  it immediately follows that the Hamiltonian is not hermitian as the superpotential is complex in nature. In other words,  Kirchhoff equations  are function of holomorphic coordinates $z=x+iy$ with the corresponding momenta 
$\partial_{z_i} = \frac{1}{2}(\partial_{x_i}-i\partial_{y_i})$ and antiholomorphic  coordinates $\bar{z}=x-iy$ with the corresponding momenta 
$\partial_{\bar{z}_i} = \frac{1}{2}(\partial_{x_i}+i\partial_{y_i})$. 
One notices that the Laughlin wave function (\ref{lf}) is a product of   the van der Monde’s determinant $\prod_{N \geq i > j \geq 1}\left( z_j-z_i \right)$ times the Gaussian  weight function
$\exp\left( -\sum_{j}  \frac{1}{4l_B^2} \vert z\vert_j ^2 \right)$.  The van der Monde’s determinant is a function of holomorphic coordinates, and the the Gaussian  weight function is a function of holomorphic  and antiholomorphic coordinates. In literature these kinds of function are studied in Segal-Bargmann space.  

It is well known in literature that the Laughlin wave function is obtained using  Landau Hamiltonian with interaction, for details readers may refer to \cite{dt}. \\
The Hamiltonian is given by
\begin{equation}
H=\frac{1}{2m}\pi\cdot\pi=\hbar\omega_B(a^\dagger a+\frac{1}{2})
\end{equation} 
where, the momentum is defined in terms of minimal coupling $\pi=P+eA$ 
and $A=-\frac{yB}{2}\hat{x}-\frac{xB}{2}\hat{y}$. Then the lowering operator is defined as
\begin{eqnarray}
a&=&\frac{1}{\sqrt{2e\hbar B}}(\pi_x-i\pi_y)\\
&=&\frac{1}{\sqrt{2e\hbar B}}\left(-i\hbar(\partial_x+\frac{yB}{2}) +i\hbar(i\partial_x-i\frac{xB}{2})\right).
\end{eqnarray}
Using the complex coordinates $z=x+iy$ and the corresponding momenta 
$\partial_{z_i} = \frac{1}{2}(\partial_{x_i}-i\partial_{y_i})$ is a holomorphic function and corresponding antiholomorphic  coordinates $\bar{z}=x-iy$ and the momenta 
$\partial_{\bar{z}_i} = \frac{1}{2}(\partial_{x_i}+i\partial_{y_i})$.
which allows us to define the following raising and lowering operators
\begin{equation}
a^{\dagger}=-i\sqrt{2}(l_B\partial_z-\frac{1}{4l_B}\bar{z}),\;\;\;\;\;\;\;\;\; a=-i\sqrt{2}(l_B\partial_{\bar{z}}+\frac{1}{4l_B}\Omega z)\label{l1}.
\end{equation} 
Then the lowest landau levels are defined as $a\psi_{LLL}=0$ where
\begin{equation}
\psi_{LLL}=f(z)\exp\left( -  \frac{1}{4l_B^2} \vert z\vert ^2 \right),
\end{equation}
where $f(z)$ is holomorphic function. For particles $N>2$ Laughlin proposed
\begin{equation}
\psi_{LLL}(z_1\cdots z_n)=f(z_1\cdots z_n) exp\left( -\sum_{i=1}^n  \frac{1}{4l_B^2} \vert z\vert_i ^2 \right),
\end{equation}
and Laughlin proposed Ground state filling fraction $\nu=1/m$ 
\begin{equation}
\psi(z_1\cdots z_n)=\prod_{i<j}(z_i-z_j)^m exp\left( -\sum_{i=1}^n  \frac{1}{4l_B^2} \vert z\vert_i ^2 \right).
\end{equation}
Here it should be noted that these raising and lowering operators defined in equation (\ref{l1}) are identical to  the intertwining operators $\hat{A}$ and 
$\hat{A}^{\dagger}$ defined in equation (\ref{eq:A}).

Then the Hamiltonian reads as
\begin{equation}
H\psi(\vert z\vert)=a^{\dagger}a \psi(\vert z\vert)=(-\partial_z+\Omega\bar{z}) (\partial_{\bar{z}}
+\Omega z)\psi(\vert z\vert),\label{ha1}
\end{equation}
where $\Omega=\frac{1}{4l_B^2}$, which gives 
\begin{eqnarray}
(\partial^2_{\vert z\vert}
+\Omega\bar{z}\partial_{\bar{z}} -\Omega z\partial_z
+\Omega^2\vert z\vert^2)\psi(\vert z\vert)=0.\label{dlu}
\end{eqnarray}
The equation (\ref{dlu}) is the complex Hermite polynomial differential equation \cite{chp} and its solutions are given by 
\begin{equation}
\psi (\vert z\vert)={\frac {1}{\sqrt {2^{n}\,n!}}}\cdot \left(\Omega\right)^{1/2}\cdot e^{-\Omega^2\vert z\vert^2}\cdot H_{n}\left(\Omega\vert z\vert\right), n=0,1,2,\ldots\label{ch}
\end{equation}
Therefore, for single particle Laughin wave function is solved using 
 Kirchhoff equations and their solutions are found to be complex Hermite polynomials.
\section{Wave Particle duality}
One of the major differences between classical mechanics and quantum mechanics is  Heisenberg uncertainty principle in terms of position and momentum given by
equation (\ref{hs}).
It is widely accepted that Heisenberg uncertainty relationship is a restatement of wave particle duality, which is the corner-stone of quantum mechanics. 
It states that quantum particles are both particles and waves. There are several theories based on wave particle duality, of which the most widely accepted one is the principle of complementarity.  The essence of this principle is that quantum systems exhibit both wave and particle nature. However, the outcome of an event is  dependent purely on an  experiment. Either the wave or the particle nature is observed but both will not be observed simultaneously in a given experiment.
Here, we address the wave particle duality through Kirchhoff equations.

Consider Kirchhoff equations (\ref{pt1}) with identical say $\Gamma$ are given by
\begin{equation}
\frac{d \bar{z}_i}{dt}=\sum_{1\leq i\leq n,i\neq j}^n\frac{i\Gamma}{z_i-z_j}+ iW(z_i).\label{pto}
\end{equation}  where $z_i=x_i+y_i$.
By using  the following identity \cite{met} and 
 substituting equation (\ref{id}) in Kirchhoff equations (\ref{pto})  one gets
\begin{equation}
f''(z_j)+\mathcal{W}(z_i)f'(z_j)= u(z_i)+iv(z_i)=0,\label{pokyl}
\end{equation} 
where $u(z_i)$ and $v(z_i)$ are real valued function.
 If a relative phase in the Kirchhoff equations (\ref{pokyl}) is developed and $v(z_i)$ is not a constant, it gives rise to wave particle duality. As an illustration, we consider the case $v(z_i)$ is constant, which gives rise to optical vortices.
 
  The  optical vortices are defined as  phase dislocation on the
beam axis, the quantised orbital angular momentum, of a Laguerre-Gaussian
laser mode \cite{allan}. Optical vortex beams are described in terms of Laguerre-Gaussian modes which are good approximation to the vortex modes created from
Hermite-Gaussian laser modes \cite{lg,lg1}. 
It is well known that the Laguerre-Gaussian vortex beam state
arises as a solution ot the paraxial approximation of the
Helmholtz equation for light or Schr\"odinger  equation for
electrons in free space. The vortex states are solutions to the
Schr\"odinger equation, Klein-Gordon equation and Dirac equations
 \cite{ev,ev1,ev2,ev3}. The connection between optical vortices and Hydrodynamiccs is studied in ref \cite{hy}.  
 
 In the equation (\ref{pokyl}), if  $v(z_i)=P$ is a constant. Then, the equation (\ref{pokyl}) is written as
 \begin{equation}
 \frac{\partial^2 u}{\partial z_i^2}+2ik\frac{\partial u}{\partial z_i}=0,\label{hz4}
 \end{equation}
 where $k=P/\Gamma$ and  the position of $N$ vortices given by $z_\alpha=x_i+iy_i$. Under paraxial wave approximation the equation (\ref{pokyl}) admit a wave equation which describes the optical vortices.
Expressing the equation (\ref{hz4}) in terms of coordinates $x_i$ and $y_i$ gives
\begin{equation}
\frac{\partial^2 u}{\partial x_i^2}+\frac{\partial^2 u}{\partial y_i^2}+2ik\frac{\partial u}{\partial z_i}=0.\label{hz3}
\end{equation}
In optics the wave equation (\ref{hz3}) is called paraxial wave equation and Gaussian beams of any beam waist $w_0$ satisfy this wave equation \cite{loy} . In paraxial wave approximation, the term $\frac{\partial^2 u}{\partial z_i^2}$ is neglected. Substituting, the term 
$\frac{\partial^2 u}{\partial z_i^2}$ back into the equation (\ref{hz3}) gives
\begin{equation}
\frac{\partial^2 u}{\partial x_i^2}+\frac{\partial^2 u}{\partial y_i^2}+\frac{\partial^2 u}{\partial z_i^2}
+2ik\frac{\partial u}{\partial z_i}=0.\label{hz1}
\end{equation}
The equation (\ref{hz1}) is derived from
the Helmholtz equation:
\begin{equation}
\nabla^2\psi(x_i,y_i,z_i)+k^2\psi(x_i,y_i,z_i)=0\label{hz}
\end{equation} 
with
\begin{equation}
\psi(x_i,y_i,z)=u(x_i,y_i,z_i)e^{ikz_i}.\label{so}
\end{equation}
Therefore,  two dimensional $n$ point vortices evolution described by Kirchhoff equations (\ref{pto}) with the identical circulation strength with the constant  imaginary background flow will admit a wave equation under paraxial wave approximation. 

 In optics, relative phase gives rise to interference. It is well known, that the particle nature of  electrons passing through Young's double slit experiment one observe interference pattern on screen which is the wave nature of electrons. It should be noted, that Kirchhoff equations (\ref{kir1}) are one dimensional and Kirchhoff equations for the vortices  (\ref{pt1}) are two dimensional. Therefore, we claim, if Kirchhoff equations in differential form (\ref{pokyl}) admits a relative phase then it gives rise to wave particle duality.
 
 \section{Discussion}
The fractional Hall effect is a topological insulator. In case of insulators, as the gap is large, it is characterized in terms of band gap.  If the insulators are connected, say from one insulator to another, without changing the band gap, such that the system always remains in the ground state,  they are called topologically equivalent insulators. In other words, when some parameter of the Hamiltonian is slowly changed adiabatically, the ground state of the system remains unchanged.  Those insulators which cannot be connected by the slowly changing Hamiltonian are called topologically inequivalent insulators. Connecting topologically equivalent insulators gives rise to a phase transition resulting in the vanishing of gap. In this gapless state, topological invariants are quantized giving rise to current. For fractional Hall effect, the Berry connection is given in terms of  equation (\ref{b1}), and the parameter $\nu$ gives topological winding numbers.

It is clear from the above how   Kirchhoff equation (\ref{kir}) is derived from one dimensional Schr\"odinger  equation. Hence, equation (\ref{kir})  and equation  (\ref{b1}) both are Kirchhoff equations for one and two dimensions respectively. Therefore, we conclude the Schr\"odinger  equation in terms of  Kirchhoff equations is a Berry connection.

In the classical Hamilton Jacobi, equation of motion is governed by the Hamiltonian and the problem is solved by continuously transforming the Hamiltonian from the initial state to the final state through canonical transformations. In quantum mechanics,  equation of motion is governed by the Schr\"odinger equation which is described by a Hamiltonian. It is well known that the   Schr\"odinger  equation is related to classical Hamilton Jacobi equation in the limit $\hbar\rightarrow 0$.  It is not possible to continuously transform Schr\"odinger equation from the initial state to the final state through a canonical transformation as it has simple poles which are quite evident when the Schr\"odinger equation is written in terms of  Kirchhoff equations (\ref{kir}).  The analogous $n$ point vortices of hydrodynamics with the fractional quantum Hall effect has allowed us to identify  the Schr\"odinger equation  as Berry connection. Thus, making the Berry connection exact gives rise to quantisation. Hence, the quantum numbers are topological invariants arising due to singularities in  Schr\"odinger equation . Therefore, we conclude that  quantisation  arises as it continuously connects the topologically inequivalent Hamiltonians in the Hilbert space. 
  
Mapping of Schr\"odinger equation to Kirchhoff equations allows us to draw the following conclusions:  The $n$ point vortices are integrable unto  three vortices \cite{aref}. If the vortices circulation strength is identical, then the $n$ point vortices, can be solved with Stieltjes electrostatic model  and their solutions are classical orthogonal polynomials \cite{cla}. The $n$ point vortices with identical circulation strength corresponds to Schr\"odinger  equation and the solutions are classical orthogonal polynomials, and they are the basis in Hilbert space. Hence, the first postulate : The state of a quantum mechanical system is completely specified by a function 
$\Psi({\bf r}, t)$ is a vector in complex Hilbert space. The complex Hilbert space arise because the Kirchhoff equations are solved using Stieltjes electrostatic model with imaginary charges \cite{kvs}. In Stieltjes electrostatic model, the $n$ moving charges between two fixed charges attain the equilibrium at the zeros of classical orthogonal polynomials depending on the position of the fixed charges. That is, if the position of fixed charges are at $\pm \infty$ the system attains equilibrium at zeros of Hermite polynomials. Hence, it is the statistical distribution of these charges is given in terms of probability distribution.  The probability distribution in terms of wave function is given by
\begin{equation}
\int_{-\infty}^{\infty} \Psi^{*}({\bf r}, t) \Psi({\bf r}, t) d\tau = 1.
\end{equation} 
These Kirchhoff equations  satisfy  a Poisson bracket relationship in phase space which is identical to the Heisenberg uncertainty relationship. Therefore, we conclude that the Kirchhoff equations, being classical equations,  describe both a particle and a wave equation nature of single particle quantum mechanics. 
 
\section{Conclusion}
In this paper, we have shown that the Kirchhoff equations are derived from  the Schr\"odinger  equation by assuming the wave function to be a polynomial like solution. These Kirchhoff equations  describe the evolution of $n$ point vortices in hydrodynamics. 
In two dimensions,  Kirchhoff equations are used to demonstrate the solution to single particle Laughlin wave function as complex Hermite polynomials. We have also shown that the equation for optical vortices, a two dimentional system, is derived from Kirchhoff equation by using  paraxial wave approximation.  These Kirchhoff equations  satisfy  a Poisson bracket relationship in phase space which is identical to the Heisenberg uncertainty relationship. Therefore, we conclude that being classical equations,  the Kirchhoff equations,  describe both a particle and a wave nature of single particle quantum mechanics in two dimensions.

\end{document}